\documentclass[12pt]{iopart}
\usepackage{bm}
\usepackage{braket}
\usepackage{graphicx}

\begin{document}

\title{Fast hybrid density-functional computations using plane-wave basis sets}

\author{Ivan Carnimeo$^{a,b}$, Stefano Baroni$^{a,c}$, Paolo Giannozzi$^{b,c}$}
\address{(a) SISSA -- Scuola Internazionale di Studi Avanzati, via Bonomea 265, 34136 -- Trieste, Italy \\ 
(b) Dipartimento di Scienze Matematiche, Informatiche e Fisiche, Universit\`a degli Studi di Udine, via delle Scienze 208, 33100 -- Udine, Italy \\ 
(c) CNR-IOM DEMOCRITOS, SISSA, Trieste, Italy}

\begin{abstract}
A new, very fast, implementation of the exact (Fock) exchange operator for electronic-structure calculations within the plane-wave pseudopotential method 
is described and carefully validated. Our method combines the recently
  proposed Adaptively Compressed Exchange approach, to reduce the number of
  times the exchange is evaluated in the self-consistent loop, with an orbital
  localization procedure that reduces the number of exchange integrals to
  be computed at each evaluation.
  The new implementation, already available in the \textsc{Quantum} ESPRESSO
  distribution, results in a speedup that is never smaller than 3-4$\times$ and
  that increases with the size of the system, according to various realistic
  benchmark calculations.
\end{abstract}
\maketitle
 
\section{Introduction}
\label{sec:intro}
Hybrid functionals, resulting from the introduction of a Fock exchange energy term into a density-functional framework, 
are very popular due to their accuracy and predictive capabilities.\cite{b88,lyp,hse}
Nowadays most if not all electronic-structure codes implement them.
In quantum chemistry codes, and in general in codes using localized basis
sets, exact-exchange terms do not significantly add to the computational
complexity of electronic states calculations.
When working with a plane-wave (PW) basis set, however, the calculation of the exact-exchange
term, although straightforward,\cite{chawla1998,boffi2016,foulkes2006,gygi1986,QE2009}
is computationally heavy.
The problem lies in the delocalized nature of orbitals and of PWs:
all pairs of canonical orbitals in the system contribute to exact
exchange.
While each contribution is quickly and effectively computed
using conventional Fast Fourier Transform (FFT) techniques, the overall
computational workload scales unfavorably with the number of electrons
in the system, being proportional to the square of the number of occupied electron
states. In practice, the usage of modern and accurate hybrid functionals
together with PW basis sets is limited to relatively small systems,
described by units cells no larger than a few dozen atoms. 

Over the years, several approaches have been proposed to extend the
range of hybrid-functional calculations with PW-based methods.
We mention, \emph{e.g.}, the reduction of the dimension of the 
density matrix \cite{guidon2010}, or of the basis set, \cite{umari2013} and
improvements in the parallelization strategies\cite{Varini:2013,Nersc:2017}. 
Another promising approach is to 
leverage a localized representation of the occupied-state manifold, 
so as to reduce the number of significant exchange integrals to be computed. 
Several proposals\cite{car2009,gygi2013,mosey2017,cervera2009,cervera2007,genovese2015,distasio2014} have been put forward, mostly using
maximally localized Wannier functions.\cite{marzari1997,marzari2012}
This approach is very effective 
and has proven to be reliable for large systems, showing that strategies based on molecular orbital localization 
are a robust route in order to cope with scaling problems in exact exchange calculations.
Furthermore, another effective method named \emph{Adaptively Compressed Exchange} (ACE) has been recently proposed by Lin Lin,\cite{linlin_ACE1}
and integrated in {\sc Quantum ES\-PRESSO}\cite{QE2017},
achieving a speed-up factor of 4-5x by the integration of a projected exchange operator in the double-loop structure of the 
Self Consistent Field (SCF) algorithm.
Notably such approach is very similar to other projection procedures previously proposed for PW calculations\cite{duchemin2010}
and also to inner-projection methods,\cite{lowdin1965,lowdin1971} popular in Quantum Chemistry.\cite{aquilante,beebe1977,koch2008,denfit}
In this work we combine for the first time an algorithm based on the molecular orbital localization
with the ACE method. 
We show how the ACE approach can be extended to deal with approximate exchange potentials
coming from the localization, and we show accurate benchmark calculations showing 
the computational performances on realistic systems.
The new method, L-ACE, allows significant 
computational gains with respect to the previous implementation of 
exact exchange with PWs.
As localization algorithm we use a modified version of the \emph{Selected Columns of the Density Matrix} (SCDM) approach,\cite{linlin_SCDM, linlin_fastSCDM, linlin_scdmk}
an algebraic technique recently introduced by Damle \emph{et al.}. 

The paper is organized as follows.
In Sec.~\ref{sec:L-ACE} we describe the L-ACE method.
Sec.~\ref{sec:methods} briefly introduces computational details
used in the extensive benchmarks of Sec.~\ref{sec:results}.
Sec. \ref{sec:conclusions} contains our conclusions.

\section{The L-ACE approach}
\label{sec:L-ACE}
In the framework of density-functional theory (DFT), 
the energy for a hybrid functional\cite{becke1993}
in a system with $N_P$ Kohn-Sham (KS) molecular orbitals 
(or pseudo-orbitals if pseudo-potentials are used), 
$\{ \psi_k(\mathbf{r}) \}_{k=1}^{N_P}$, 
can be written in terms of the charge density,
$n (\mathbf{r}) = \sum_i^N \; | \psi_i (\mathbf{r})|^2$, 
and of the density matrix,
$\gamma (\mathbf{r},\mathbf{r}\,') = \sum_i^N \; \psi_i(\mathbf{r}) \psi^*_i(\mathbf{r}\,')$,
where $N$ is the number of occupied orbitals.
In the following the index $i$ will be used to label the occupied orbitals ($i=1,..,N$), 
the indexes $j$ and $k$ will be used for the KS orbitals ($j,k=1,...,N_P$), 
also including virtual orbitals if $N_P>N$. 
The spin part, omitted for simplicity of notations, should be assumed wherever relevant.
In order to determine the molecular orbitals,
a set of single-particle KS equations
\begin{equation}
\bigg( -\frac{1}{2}\nabla^2 + \hat{V}_{ext} + \hat{V}_{Hxc} [n, \alpha] 
                           - \alpha \hat{V_X} [\gamma] \bigg) \psi_k(\mathbf{r}) = \varepsilon_k \psi_k(\mathbf{r})
\end{equation}
must be solved.
The first term of the last equation is the kinetic energy operator, 
the second term is the external potential, 
$\hat{V}_{Hxc}$ includes the Hartree and the DFT exchange-correlation contributions, 
and $\hat{V}_X$ is the exact (Fock) exchange operator
(with a minus sign with respect to the usual definition, to make it
positive definite).
The empirical parameter $\alpha$ is used to weight the Fock and DFT exchange contributions, 
and takes different values according to the particular functional parametrization 
(e.g. see Refs.~\cite{b3lyp,pbe0,hse}).
The Fock exchange operator can be defined through its action on a generic function $\psi_k$ 
\begin{equation}
\label{eq:V}
\hat{V}_X\psi_k (\mathbf{r}) = e^2 \sum_{i=1}^N  \psi_i (\mathbf{r}) \int d\mathbf{r}\,' \frac{\psi_i^* (\mathbf{r}\,') \psi_k (\mathbf{r}\,')}{|\mathbf{r}-\mathbf{r}\,'|},
\end{equation}
$e$ being the electronic charge.
When a PW basis set is used, Eq.~\ref{eq:V} is usually represented in reciprocal space as a matrix $\mathbf{V}_X$ 
of dimension $N_P\times N_{PW}$, where $N_{PW}$ is the number of plane waves, and the integrals are numerically solved using FFT algorithms.
The latter is the bottleneck of the calculation especially when large molecular systems are involved, 
because many FFTs must be performed over large grids.
A significant speed-up can be achieved
by employing the ACE method.\cite{linlin_ACE1} 
In the ACE method $\mathbf{V}_X$ is used to build the ACE operator, 
which can be written in matrix form as
\begin{equation}
\label{eq:ACE1}
\mathbf{W}_X = \mathbf{V}_X \cdot \mathbf{M}^{-1} \cdot \mathbf{V}_X^\dagger 
\end{equation}
where $\mathbf{M}_{jk}=\braket{\psi_j|\hat{V}_X|\psi_k}$ is the exchange matrix.
In practice the $\mathbf{M}^{-1}$ matrix is factorized via the Cholesky decomposition and the 
ACE operator takes the form
\begin{equation}
  \mathbf{W}_X = \bm{\xi}\cdot\bm{\xi}^\dagger
\end{equation}
where
\begin{equation}
  \bm{\xi} = \mathbf{V}_X \cdot \mathbf{L}^{-T}
\end{equation}
and $\mathbf{L}^{-T}$ is the Cholesky factor of $\mathbf{M}^{-1}$.
The application of the ACE operator to a generic function
costs as little as $N_P$ scalar products and it is much cheaper than the
application of $\hat{V}_X$. 
When a double-loop SCF algorithm is employed, 
as in the \texttt{pw.x} code of {\sc Quantum ESPRESSO},\cite{QE2009,QE2017}
it is much more convenient to compute $\mathbf{V}_X$ only to build the ACE operator, 
using the latter to solve the KS equations iteratively,
instead of computing $\mathbf{V}_X$ at every SCF iteration.
Once convergence is achieved, $\mathbf{V}_X$ is evaluated again in order to update the ACE operator, 
the KS equations are solved with the new operator, 
and this process is repeated until the total energy reaches the minimum.
As a result, the exact exchange potential needs not to be evaluated inside every SCF iteration, 
but only once per SCF loop, resulting in a significant speed-up.
It is important to notice that the ACE operator is perfectly
equivalent to $\hat{V}_X$ only at convergence or whenever it
is applied to the same set of functions used to build it.
In all other cases, it is an approximation, whose accuracy
improves as more functions are used for the projection, 
and it can been shown that its expectation values 
are greater than the expectation values of $\hat{V}_X$.\cite{linlin_convergence, lowdin1965, lowdin1971}
The ACE method has been implemented in the {\sc Quantum ES\-PRESSO} code\cite{QE2017} with minor modifications,
so that we refer to the original work\cite{linlin_ACE1} for a detailed description of the method.

In this work we have achieved a further speed-up in the exact exchange evaluation, 
by projecting the ACE operator on a set of localized orbitals, $\{ w_i \}_{i=1}^N$ 
and neglecting the integrals of Eq.~\ref{eq:V} involving two orbitals localized in different regions of space.
Among the many known localization schemes,\cite{car2009,gygi2013,mosey2017,marzari1997,marzari2012,parrinello1999,vonniessen1972,linlin_SCDM,linlin_fastSCDM,linlin_scdmk} 
we have chosen the \emph{Selected Columns of the Density Matrix} (SCDM) approach,\cite{linlin_SCDM,linlin_fastSCDM}
which has the advantage of being fast, non-iterative,
and to be estensible to periodic systems using k-points\cite{linlin_scdmk}.
The method proposed in the following is however completely general
and can be applied independently of the particular choice of the localization algorithm. 

In order to decide whether a given exchange integral can be discarded or must be retained, 
a reliable criterion needs to be found.
In the limit case of orbitals shaped like Gaussian functions,
the exchange integrals 
depend linearly upon the overlap of the two functions, 
and can be thus neglected when the overlap is small enough.
Extending the idea to more general functions which can take positive and negative values, 
a safe criterion is that whenever the overlap of the moduli
\begin{equation} 
\label{eq:Sij}
S_{ik} = \int d\mathbf{r} \quad |w_i(\mathbf{r})| \cdot |w_k(\mathbf{r})|  
\end{equation}
is smaller than a predefined threshold ($S_{thr}$), i.e. $S_{ik} < S_{thr}$,
the corresponding exchange integral can be set to zero. 

The (approximate) matrix of truncated exchange potential, $\mathbf{V'}_X$, 
can be formally written as 
\begin{equation}
  \label{eq:pot}
\mathbf{V'}_X = \mathbf{V}_X -  \sum_{\alpha\beta \atop S_{\alpha\beta}<S_{thr}} \mathbf{u}^{(\alpha\beta)}
\end{equation}
where $\alpha$ and $\beta$ run over the discarded pairs of orbitals, 
and $\mathbf{u}^{(\alpha\beta)}$ have all columns set to zero except two:
\begin{equation}
\mathbf{u}^{(\alpha\beta)}_\alpha (\mathbf{r}) = e^2 w_\beta(\mathbf{r})  \int d\mathbf{r}\,' \frac{ w^*_\beta(\mathbf{r}\,')  w_\alpha(\mathbf{r}\,') } {|\mathbf{r}-\mathbf{r}\,'|} 
\end{equation}
and $\mathbf{u}^{(\beta\alpha)}_\beta (\mathbf{r})$, with switched $\alpha$ and $\beta$ indexes. 
The truncated exchange matrix $\mathbf{M'}$ can then be written as 
\begin{equation}
\label{eq:error}
\mathbf{M'} = \mathbf{M} - \sum_{\alpha\beta \atop S_{\alpha\beta}<S_{thr}} \mathbf{m}^{(\alpha\beta)},
\end{equation}
where $\mathbf{m}^{(\alpha\beta)}$ contains only column vector corrections.
The matrix $\mathbf{M'}$ is still invertible and the matrix inversion lemma (Woodbury identity)\cite{henderson1981,miller1981} allows to write 
$\mathbf{M'}^{-1} = \mathbf{M}^{-1}  + \Delta\mathbf{M}$.
Furthermore, recalling that the determinant is a linear function of the
columns (rows) 
and that the $\alpha$ and $\beta$ columns of $\mathbf{m}^{(\alpha\beta)}$ are small by construction,
we can safely assume that 
$\textrm{det}\left( \mathbf{M'} \right) \simeq \textrm{det} \left( \mathbf{M} \right),$
where $\textrm{det}\left(\mathbf{M}\right) >0$
is the product of positive definite eigenvalues.
This suggests that the truncated exchange matrix $\mathbf{M'}$
  remains positive definite and invertible, provided that the
  $\mathbf{m}^{(\alpha\beta)}$ matrices are small compared with $\mathbf{M}$.
The approximate ACE operator then reads 
\begin{equation}
\label{eq:ACE3}
\mathbf{W'}_X = \mathbf{V'}_X \cdot \mathbf{M'}^{-1} \cdot \mathbf{V'}_X^\dagger
             \simeq \mathbf{W}_X+ \Delta \mathbf{W}.
\end{equation}

The last equality just formalizes the fact that the approximate ACE operator can be written as the sum 
of the exact ACE operator plus a truncation error.
The ACE method exploiting the localization algorithm will be referred to as the Localized ACE (L-ACE) method in the following.
Unlike $\mathbf{M}$, the $\mathbf{M'}$ matrix is
not symmetric and it has to be symmetrized in order to perform the Cholesky
decomposition. We have tested three different symmetrizations: 
copying the upper (lower) triangle into the lower (upper) triangle, 
and replacing the matrix with $(\mathbf{M'}+\mathbf{M'}^T)/2$.
In all cases, also after the symmetrization the matrix can be decomposed
as in
Eq.~\ref{eq:error}, into a sum of column-wise (or row-wise) corrections
and the method is still valid.

One drawback of the L-ACE method is that the truncation introduces a small error
in the exchange energy, 
resulting in an oscillatory behaviour of the total energy during the self-consistency cycle
which may plague the convergence,
as observed in other linear-scaling
approaches based on truncated potentials.\cite{cervera2007,cervera2009}
However, in our case the truncation is done only on the exact
exchange energy, which is a small fraction of the total energy, 
so that such fluctuations are very small.
Furthermore, such fluctuations depend on the threshold $S_{thr}$, 
which can be tuned to be large enough to allow 
significant computational savings but still small enough to allow smooth SCF convergence. 
For the molecular systems reported in Section~\ref{sec:results} we have verified that 
values of the threshold in the interval 
$0 < S_{thr} \leq 0.005$ allow to achieve a significant speed-up
and yet converge smoothly the SCF to an estimated accuracy of $10^{-6}$ Ry.
When very tight convergences are required, 
a good strategy to improve SCF convergence is to align 
the orbitals at a given iteration to the ones at the previous step,
thus ensuring convergence to arbitrary precision.

\section{Computational methods}
\label{sec:methods}
All calculations have been performed using a development version
\footnote{to be released by the time this article is published;
  work is currently under way to port the same methodology to the GPU
  version of {\sc Quantum ESPRESSO}.}
of the  
{\sc Quantum ESPRESSO} \texttt{pw.x} code.\cite{QE2009,QE2017}
Pure DFT calculations have been performed using the PBE\cite{pbe} functional, 
while PBE0,\cite{pbe0} B3LYP,\cite{b3lyp} and HSE\cite{hse} functionals have been used for hybrid calculations. 
Three FFT grids are defined for hybrid functional calculations, 
one for KS orbitals, one for the charge density and one for the exchange potential. 
Such grids are uniquely identified by the cell parameters and the cutoff values, 
which will be indicated in the following for each test system. 
Norm-conserving pseudopotentials have been used throughout the work.
We will refer to the old method for the exact exchange,
not using either ACE or localization, as the "Full" method;
to the ACE algorithm without localization as "ACE";
to ACE plus localization as "L-ACE($S_{thr}$)'', with the value
of $S_{thr}$ always specified (e.g., L-ACE(0.004) means that we are
neglecting all exchange integrals in the potential with $S_{ik}<0.004$).

The calculations have been run on the Galileo supercomputer hosted at CINECA. 
In the current configuration, the nodes of the system are based on a IBM NeXtScale architecture 
equipped with 2 x 18-cores Intel Xeon E5-2697 v4 (Broadwell) running at 2.30 GHz nominal clock speed and 128 GB of RAM memory. 
The nodes are connected through a QDR InfiniBand network with 40 Gbps links.

\section{Results and discussion}
\label{sec:results}
\begin{figure}[!ht]
    \centering
    \parbox{0.40\textwidth}{
      \includegraphics[width=0.25\textwidth]{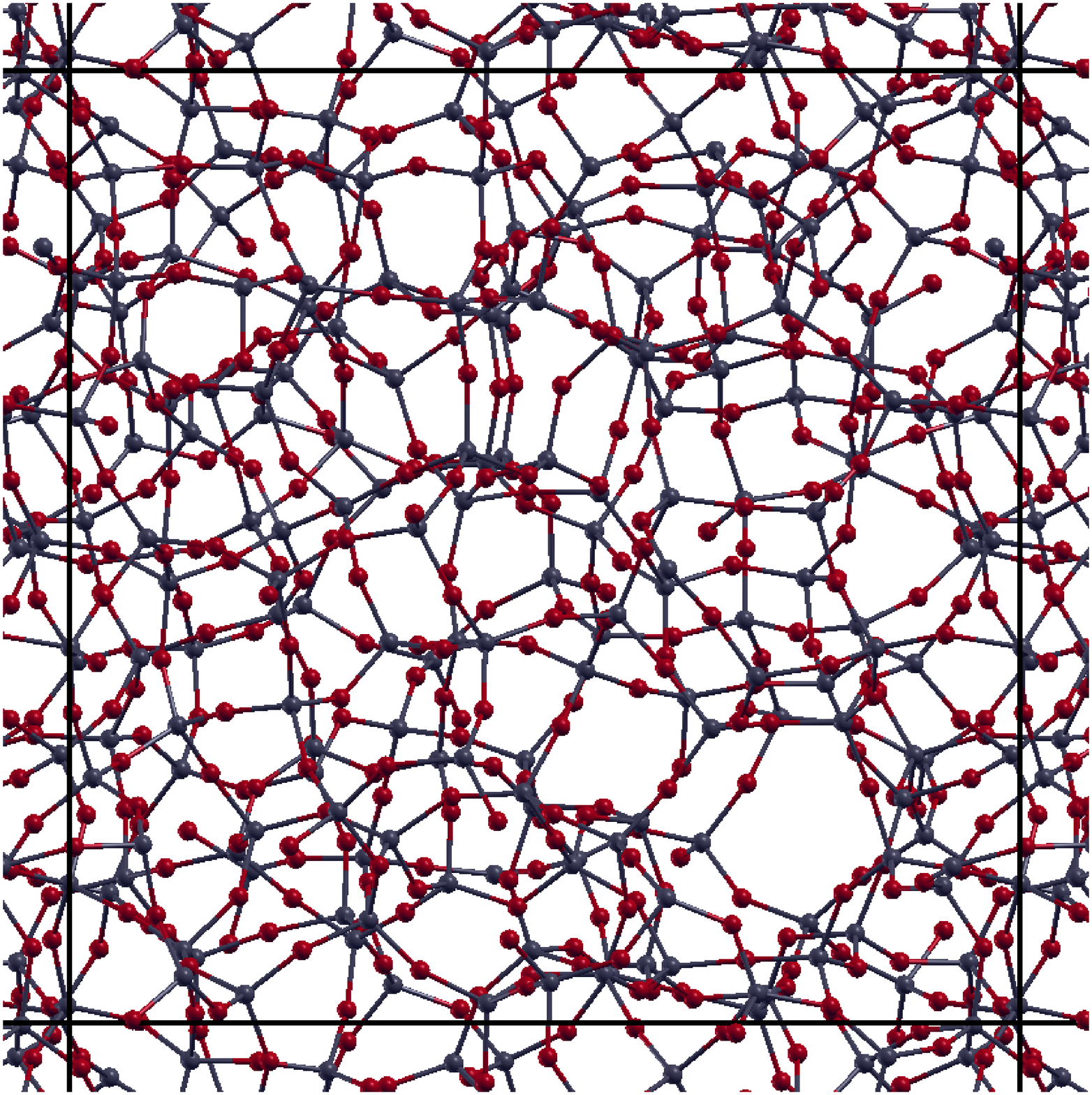} \\
        {\footnotesize (a) Amorphous Silica (576 atoms).}
    }
    \centering
    \parbox{0.40\textwidth}{
      \includegraphics[width=0.25\textwidth]{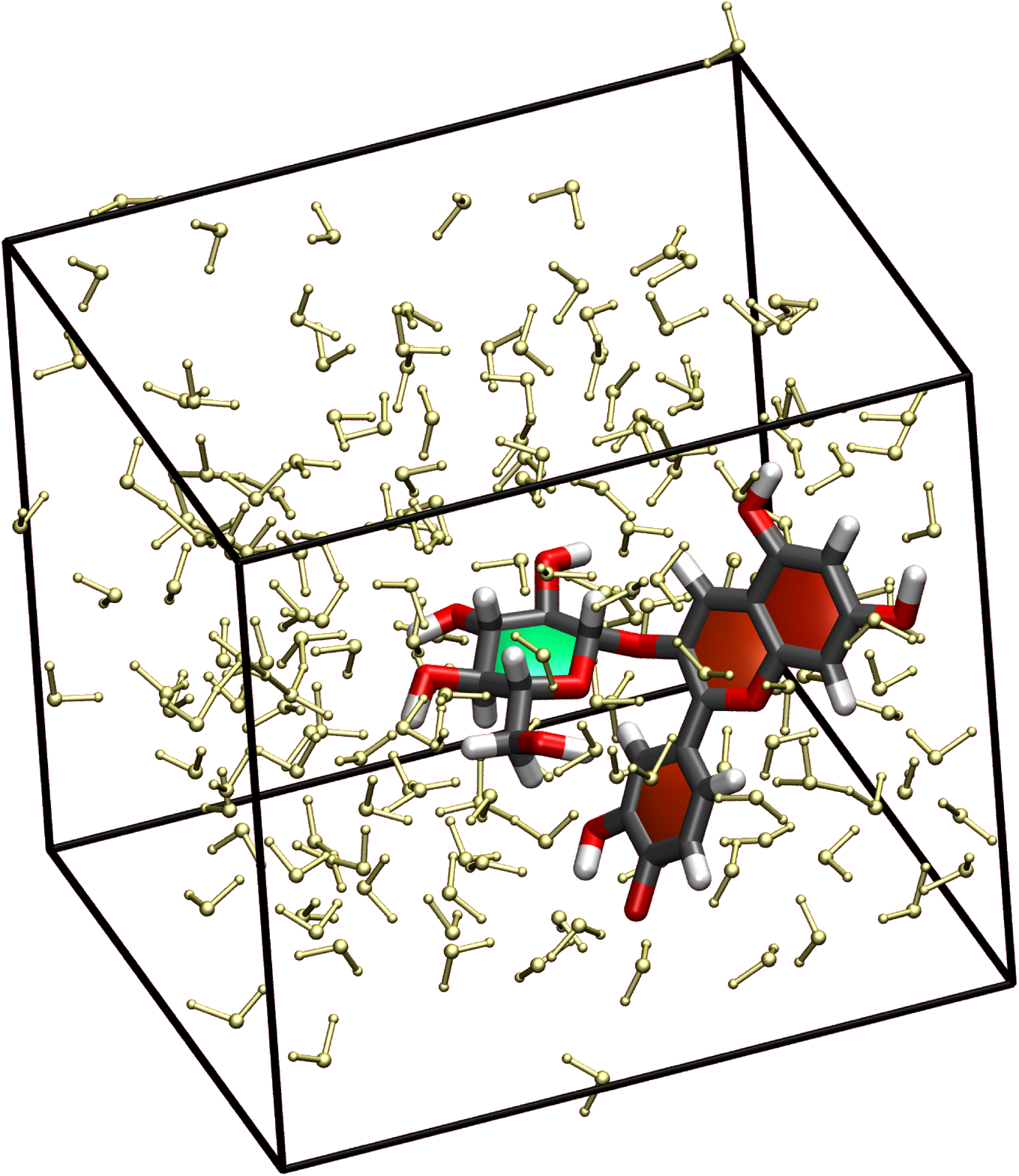} \\
        {\footnotesize (b) C3OG in water (565 atoms).}
    }
    \centering
    \parbox{0.40\textwidth}{
      \includegraphics[width=0.25\textwidth]{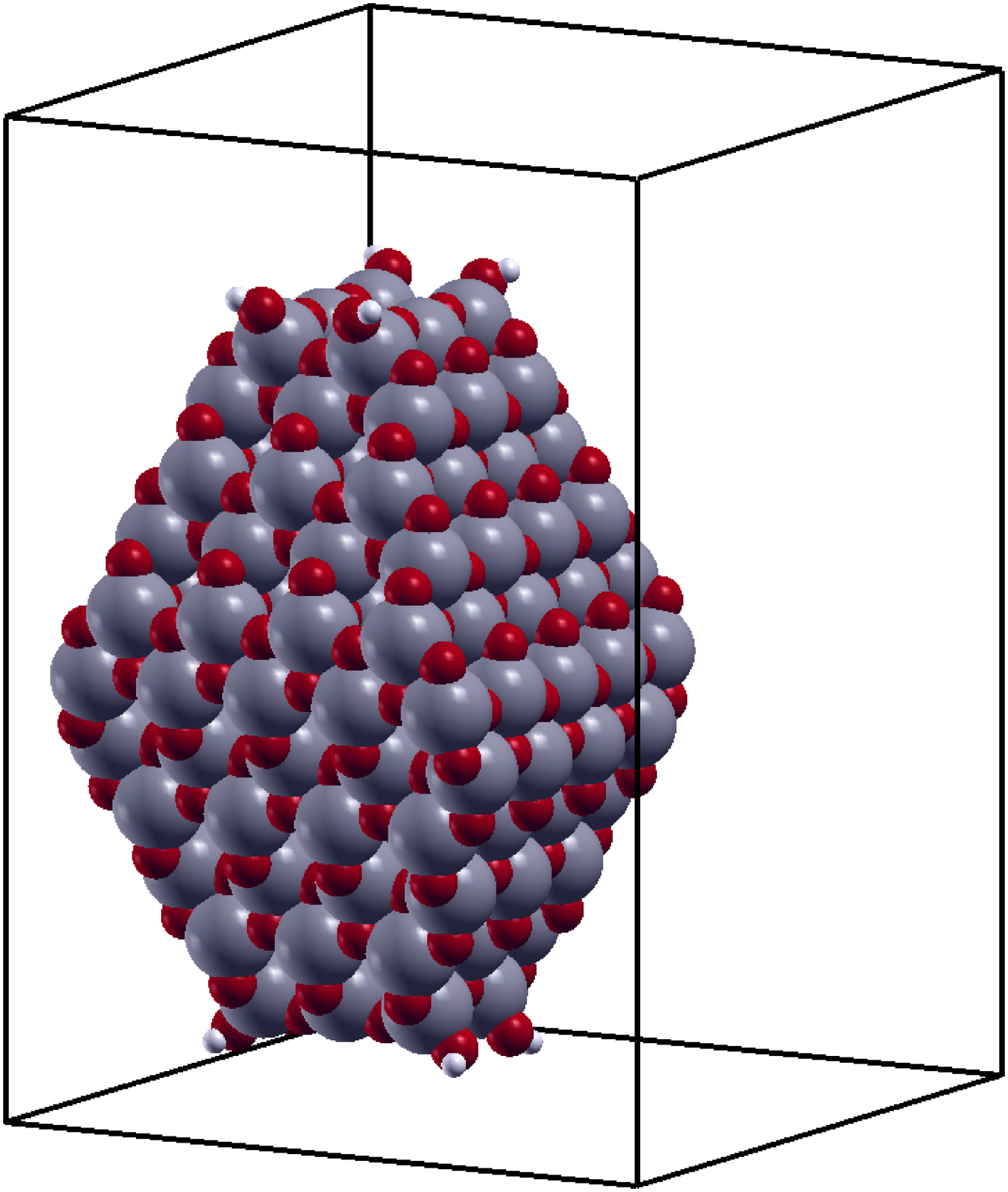} \\
        {\footnotesize (c) TiO$_2$ cluster (465 atoms).}
    }
\caption{ \label{fig:all} Unit cells of three representative systems.
  \footnotesize }
\end{figure}

We used three different typologies of molecular systems
(shown in Figure~\ref{fig:all})
to validate the truncation scheme of the L-ACE method,
ranging from amorphous materials (Silica),
to organic molecules (Anthocyanine in water)
and nanoparticles (TiO$_2$ cluster). 

\subsection{Amorphous Silica}
Five different amorphous Silica (SiO$_2$) geometries, containing 72, 144, 288, 432 and 576 atoms,
have been generated by classical MD simulations using the BKS force field,\cite{bks_1,bks_2}  
quenching from the melt at 6500K down to 500K at a constant quenching rate, as described in Ref.\cite{ercole2017}
The geometries are in cubic cells of dimension 18.6826, 24.1996, 30.3976, 35.1143 and 38.6458 Bohr, respectively.
In Figure~\ref{fig:all}a the largest system of 576 atoms is shown.
We have used such geometries for hybrid SCF calculations, 
using the PBE0 functional, and cutoffs of 80 Ry for the orbital grid
and 320 Ry for the density and the exchange grid.
For pure DFT calculations the PBE functional has been used.

\begin{figure}[!ht]
\centering
\includegraphics[width=0.5\textwidth]{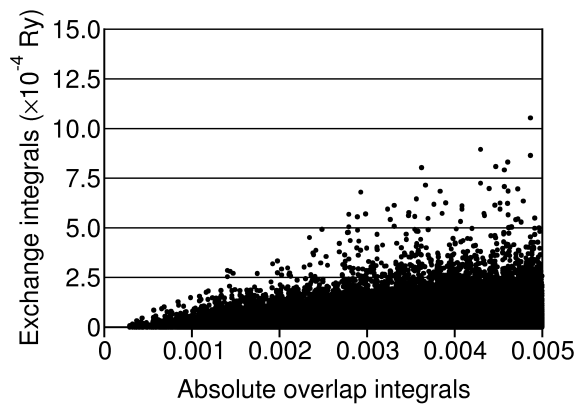}
\caption{ \label{fig:histogram}
\footnotesize 
Exchange integrals of type $(jiik)$ and $(jkki)$ plotted versus the corresponding 
absolute overlap integrals $S_{ik}$, for any $j$.
}
\end{figure}

First of all we have numerically checked the consistency of the L-ACE approximations.
In practice the L-ACE method is based on the assumption that 
whenever the absolute overlap $S_{ik}$ is small,
all the exchange 
integrals\footnote{In this notation\cite{szabobook}  for the electrostatic
  integrals, the first (last) two indexes refer to functions integrated
  over $\mathbf{r}$ ($\mathbf{r}\,'$)}
of the type $(jkki)$ and $(jiik)$ are also small for any $j$.
In Figure~\ref{fig:histogram}, 
the values of $(jkki)$ and $(jiik)$ integrals are plotted versus the corresponding values of $S_{ik}$, 
for the 72-atoms system.
We observe that the largest integrals
-- those of the type $(kkki)$ and $(iiik)$ --  
show a linear dependence upon $S_{ik}$ in the range between 0 and 0.005,
confirming that the absolute overlap is a good quantity for the prescreening.
Following Figure~\ref{fig:histogram}, values of $S_{thr}$ in the range
between 0 and 0.005 have been used for all the calculations in this work.

\begin{figure}[!ht]
\centering
\includegraphics[width=0.5\textwidth]{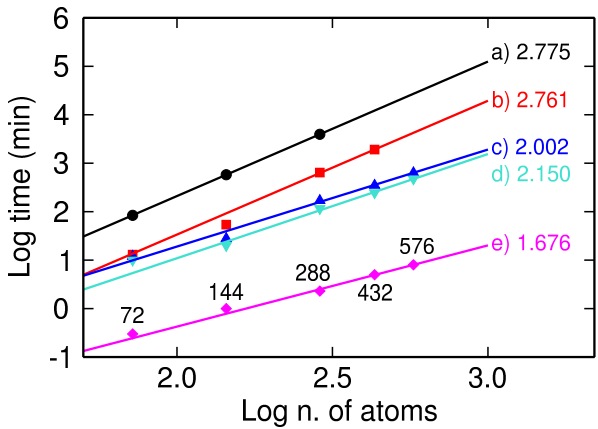}
\caption{ \label{fig:timesize}
\footnotesize 
Wall time for a complete SCF (logarithmic scale) in Silica
versus the logarithm of the number of atoms.
a) Full, b) ACE, c) L-ACE(0.002), d) L-ACE(0.004), e) DFT.
DFT calculations are with the PBE functional, all the others are with the PBE0 functional.  
The lines are the plot in logarithmic scale of the fitting function $y=\alpha x^\beta$ and $\beta$ is reported for each method.
The number of atoms is shown for clarity near the magenta (e) line.
}
\end{figure}

In Figure~\ref{fig:timesize} the WALL times for one SCF with standard convergence 
($10^{-6}$ Ry) are plotted as a function of the number of atoms
of the Silica cells in a logarithmic scale, in order to visualize the scaling with the system size. 
The same numbers are shown in the last column of Table~\ref{tab:silica}.
When the Full algorithm is used, we could perform the SCF 
only for the three smallest systems, 
and the SCF for the system of 288 atom took nearly three days. 
The use of the ACE method improves the performance, 
and we could perform calculations also for the system composed by 432 atoms, 
in a reasonable time.
With the L-ACE approach, the WALL times are strongly reduced especially 
for the largest systems, 
and the calculation of the system composed by 576 atoms becomes feasible.
Noteworthly, in all cases PBE0 calculations are still much slower than
pure PBE ones.
The lines in Figure~\ref{fig:timesize} are fits of the computational times to a function of the type
$y = \alpha x^\beta$, where $y$ is the SCF time and $x$ the number of atoms composing the system,
and the values of $\beta$ are reported in the Figure.
We observe that although the Full method is much more expensive than ACE, they scale pretty much in the same way with increasing system size:
$\beta\simeq 2.8$ in both cases.
For comparison, L-ACE scales with an exponent of about $2$,
closer to the pure DFT value of about $1.7$. 
The behavior of the L-ACE(0.002) and L-ACE(0.004) cases can be explained by
observing (Table~\ref{tab:silica}) that the percentage of integrals included
in the calculation is 91\% and 71\% respectively for the 72-atom system,
12\% and 8\% respectively for the largest system.

\begin{figure}[!ht]
\centering
\includegraphics[width=0.5\textwidth]{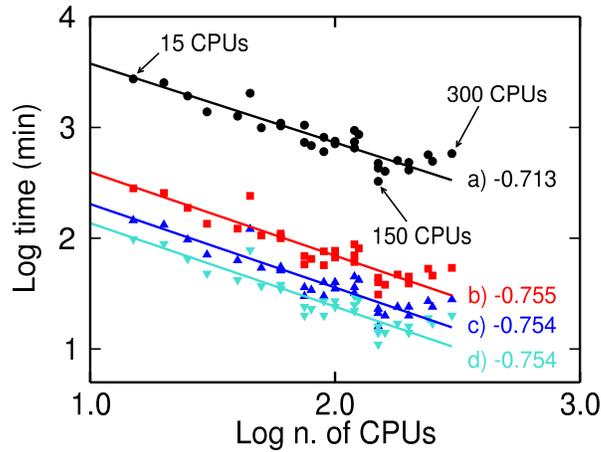}
\caption{ \label{fig:timecpu}
\footnotesize 
Wall time for a complete SCF (logarithmic scale) versus the logarithm of the number of CPUs used in the calculations,
for the 144-atom Silica cell.
a) Full, b) ACE, c) L-ACE(0.002), d) L-ACE(0.004).
With pure DFT method SCF times were always lower than 3 min. 
The lines are the plot in logarithmic scale of the fitting function $y=\alpha x^\beta$ and $\beta$ is reported for each method.
}
\end{figure}

\begin{figure}[!ht]
\centering
\includegraphics[width=0.5\textwidth]{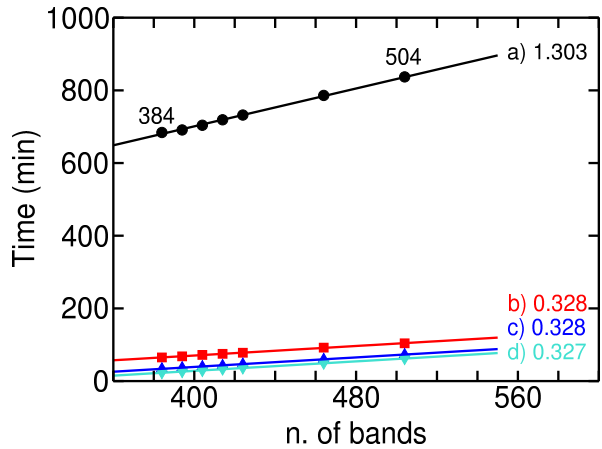}
\caption{ \label{fig:timebands}
\footnotesize 
Wall time for a complete SCF versus the number of bands required in the calculations,
for the 144-atom Silica cell (384 occupied states).
a) Full, b) ACE, c) L-ACE(0.002), d) L-ACE(0.004).
With pure DFT method SCF times were always lower than 1 min. 
The lines are the linear regression of the points and the slopes are reported for each method.
}
\end{figure}

It is then reasonable that the computational times of the two methods become more similar for increasing system size.
\begin{table}[h!]
\footnotesize
\begin{tabular}{llllllll}
 Atoms & Method       & $\Delta E_\textrm{hyb}$ & Max      & MAD     & St. Dev. & $\Delta \varepsilon_\textrm{HL}$ & Time         \\\hline
 72    & Full         & 0.4059                &  0.01449 & 0.00398 & 0.00519  & 7.6355                         & 84   (100\%) \\
       & ACE          & 0.4059                &  0.01450 & 0.00398 & 0.00519  & 7.6356                         & 13   (100\%) \\
       & L-ACE(0.002) & 0.4064                &  0.01450 & 0.00398 & 0.00520  & 7.6355                         & 12   (91\%)  \\
       & L-ACE(0.004) & 0.4106                &  0.01451 & 0.00399 & 0.00520  & 7.6360                         & 10   (71\%)  \\
       & pure DFT     & 0                     &  0       & 0       & 0        & 5.1161                         & 0.3  (0\%)   \\
 144   & Full         & 0.5567                &  0.01772 & 0.00404 & 0.00520  & 7.2559                         & 578  (100\%) \\
       & ACE          & 0.5567                &  0.01773 & 0.00404 & 0.00520  & 7.2560                         & 54   (100\%) \\
       & L-ACE(0.002) & 0.5592                &  0.01773 & 0.00404 & 0.00520  & 7.2562                         & 28   (51\%)  \\
       & L-ACE(0.004) & 0.5663                &  0.01774 & 0.00404 & 0.00520  & 7.2576                         & 20   (33\%)  \\
       & pure DFT     & 0                     &  0       & 0       & 0        & 4.7764                         & 1    (0\%)   \\
 288   & Full         & 0.5588                &  0.01735 & 0.00400 & 0.00520  & 7.0843                         & 3956 (100\%) \\
       & ACE          & 0.5588                &  0.01735 & 0.00400 & 0.00520  & 7.1154                         & 641  (100\%) \\
       & L-ACE(0.002) & 0.5588                &  0.01735 & 0.00400 & 0.00520  & 7.1155                         & 170  (25\%)  \\
       & L-ACE(0.004) & 0.5694                &  0.01736 & 0.00401 & 0.00520  & 7.1156                         & 116  (16\%)  \\
       & pure DFT     & 0                     &  0       & 0       & 0        & 4.6115                         & 2.3  (0\%)   \\
 432   & ACE          & 0.5772                & -0.02067 & 0.00405 & 0.00531  & 6.7921                         & 1913 (100\%) \\
       & L-ACE(0.002) & 0.5806                & -0.02068 & 0.00405 & 0.00532  & 6.7963                         & 350  (16\%)  \\
       & L-ACE(0.004) & 0.5878                & -0.02069 & 0.00406 & 0.00532  & 6.7973                         & 246  (11\%)  \\
       & pure DFT     & 0                     &  0       & 0       & 0        & 4.2792                         & 5    (0\%)   \\
 576   & L-ACE(0.002) & 0.5595                &  0.01812 & 0.00393 & 0.00507  & 7.1319                         & 636  (12\%)  \\
       & L-ACE(0.004) & 0.5666                &  0.01815 & 0.00394 & 0.00507  & 7.1330                         & 475  (8\%)   \\
       & pure DFT     & 0                     &  0       & 0       & 0        & 4.6377                         & 8    (0\%)   \\
\end{tabular}
\caption{\label{tab:silica}
  Summary of results for Silica cells. Pure DFT calculations with PBE,
  hybrid calculations with PBE0. $\Delta E_\textrm{hyb}$ is the difference
  between PBE0 and PBE total energies per atom (kcal/mol),
  $\Delta \varepsilon_\textrm{HL}$ is the HOMO-LUMO gap (eV),
  Max, MAD and St. Dev. are the maximum difference, the mean absolute
  difference, and the standard deviation, respectively, of the PBE0 force
  components (Ry/Bohr) with respect to the PBE ones,  
  Time is the WALL time (min) for one SCF with 300 CPUs,
  in parenthesis the fraction of included integrals is reported.
}
\end{table}

In Figure~\ref{fig:timecpu} the scaling with the number of processors is shown for the 144-atom system.
In order to account for the unavoidable fluctuations in the computational load of the cluster, 
and for the different memory allocations when different number of nodes are chosen, 32 runs (on 1, 2, 3, 4, 5, 6, 8 and 10 nodes,
using 15, 20, 25 and 30 cores per node) have been performed.
All the WALL times are shown in Figure~\ref{fig:timecpu} as points. 
With all the methods the lowest computational times are achieved 
when using 150 cores (marked by an arrow for the black circles in the plot).
The increase observed for more than 150 cores is due to limitations of the
default FFT parallelization scheme used by {\sc Quantum ESPRESSO}\cite{QE2009,QE2017},
that becomes ineffective when the number of processors exceeds the number
of FFT planes (144 in this case).
Among the three different combinations of cores per nodes resulting in 150 CPUs,
i.e. 10 nodes with 15 cores per node, 6 nodes with 25 cores per node and 5 nodes with 30 cores per nodes,
the best performances are obtained using 10 nodes with 15 cores per node for all methods.
Also in this case the points have been fitted with the function $y = \alpha x^\beta$, where $y$ is the SCF time
and $x$ the total number of cores, in the range 15--150 cores, 
and slopes of about -0.7 have been found for all the methods, 
with slightly better performances for ACE and L-ACE with respect to the Full method.

Finally since the exchange potential is applied to computed (occupied or empty)
Kohn-Sham orbitals, the computational times grow with the number of such
orbitals, as shown in Figure~\ref{fig:timebands}. 

In Table~\ref{tab:silica} we report the results for selected physical quantities.
Differences between PBE0 and PBE results are reported in order to have a
unique reference, since PBE calculations are available for all systems.
In the third column of Table~\ref{tab:silica} hybrid energy corrections
(that is: differences between the PBE0 and PBE total energies per atom)
are reported. The two exact methods, Full and ACE,
give an hybrid energy correction of about 0.4 kcal/mol per atom for the 72 atoms
system, of about 0.6 kcal/mol for all other systems. Results from L-ACE(0.002)
and L-ACE(0.004) differ on the third and second significant digit, respectively,
from exact results.

Columns 4 to 6 of Table~\ref{tab:silica} contain the Mean Absolute Differences
(MAD), maximum (Max) differences and standard deviation of the differences
between PBE0 and PBE forces, averaged over all the force components of all
atoms. Having converged the SCF to $10^{-6}$ Ry we can trust the forces up
to $10^{-4}$ Ry/Bohr, and five decimal digits are reported.

The maximum differences between the PBE0 force components computed with the
Full and ACE method and the PBE ones in the system of 72 atoms is 0.01449 Ry/Bohr,
while for the systems with 144 and 288 atoms they are 0.01772 and 0.01735 Ry/Bohr,
respectively. The loss of accuracy introduced by L-ACE turns out to be very
small: in all cases the maximum differences are almost exactly the same as
for ACE and occur for the same force component.

A similar analysis can be performed for the MAD and the standard deviation. 
In all cases the differences between the ACE forces and L-ACE ones 
fall on the fifth decimal digit. This suggest that the truncation scheme used
in L-ACE for the potential is reliable enough to provide good forces, that can
be used for geometry optimizations and likely for molecular dynamics as well.

While energies and forces can be obtained from occupied orbitals only,
other quantities such as e.g. the band gaps also require virtual orbitals.
The band gaps for the Silica systems reported in Table~\ref{tab:silica} 
have been calculated by adding ten virtual orbitals to the number of occupied orbitals.
Since the ACE operator is perfectly equivalent to the exact exchange operator
only when it is applied to a function belonging to the space in which the projection is valid,
$N_P$ in Eq.~\ref{tab:silica} has been set as the number of occupied orbitals 
plus ten virtual orbitals.
In the L-ACE scheme the occupied manifold is localized and the truncation scheme is applied, 
while the virtual orbitals are just those obtained from the KS equations,
and all the exchange integrals involving at least one virtual orbital are evaluated.
For this reason the speed-up of L-ACE with respect to ACE decreases
when virtual orbitals are included.
In all cases the errors in the band gaps (sixth column of Table~\ref{tab:silica}) are 
of the order of $10^{-3}$ eV, which is accurate enough for any application.

\subsection{Anthocyanine in water}
As an example of organic molecules, the Cyanidin 3-O-glucoside (C$_{21}$H$_{20}$O$_{11}$) molecule (shortcut C3OG) has been considered.
We extracted one snapshot from a Car-Parrinello simulation of the C3OG molecule surrounded by 171 water molecules,
in an orthorhombic cell with parameters $a=33.9933$ Bohr, $b=36.7973$ Bohr and $c=31.0310$ Bohr. 
The molecule is in the neutral state, the water molecules form various hydrogen bonds with the --OH moieties,
and the whole system is shown in Figure~\ref{fig:all}b, 
with the aromatic and sugar rings of C3OG highlighted in orange and green, respectively.

On such geometry we run SCF calculations using the B3LYP\cite{b3lyp} functional
and cutoffs of 80 Ry, 320 Ry, 160 Ry for the orbital, density and exchange
grids, respectively. For pure DFT calculations the PBE functional has been used.
Calculations have been performed on the C3OG molecule both in vacuum and surrounded by the water molecules.
The calculations in vacuum were performed on the same geometry as in water just removing the solvent molecules. 
This allowed to compute the solvation energies as the differences between the total energy of the molecule in water
and the total energies of the molecule and the waters alone,  
and the HOMO-LUMO band gaps in vacuum and in water.
Such data are reported in Table~\ref{tab:molecule}, 
together with the computational times and with the hybrid energies corrections, 
analogously to the Silica systems (see previous paragraph).

\begin{table}[h!]
\footnotesize
\begin{tabular}{lllllll}
 Method        & $\Delta E_\textrm{hyb}$ & $\Delta E_S$ & $\Delta \varepsilon_\textrm{HL}^\textrm{gas}$ & $\Delta \varepsilon_\textrm{HL}^\textrm{water}$ & Time        \\\hline
 ACE           & -24.9794              & -161.3846    & 2.0521                                    & 2.6091                                      & 249 (100\%) \\
 L-ACE (0.001) & -24.9790              & -161.3457    & 2.0521                                    & 2.6089                                      & 92  (36\%)  \\
 L-ACE (0.002) & -24.9780              & -161.2540    & 2.0521                                    & 2.6085                                      & 71  (26\%)  \\
 L-ACE (0.003) & -24.9767              & -161.1331    & 2.0521                                    & 2.6080                                      & 62  (21\%)  \\
 L-ACE (0.004) & -24.9750              & -161.0085    & 2.0521                                    & 2.6075                                      & 56  (18\%)  \\
 L-ACE (0.005) & -24.9730              & -160.8220    & 2.0521                                    & 2.6071                                      & 52  (16\%)  \\
 pure DFT      & 0                     & -169.4025    & 1.1085                                    & 1.5666                                      & 5   (0\%)   \\
\end{tabular}
\caption{\label{tab:molecule}
  Summary of results for C3OG molecule in water (565 atoms).
  Pure DFT calculations with PBE, hybrid calculations with B3LYP.  
  $\Delta E_\textrm{hyb}$ is the difference between B3LYP and PBE total
  energies per atom (kcal/mol), $\Delta E_S$ is the solvation energy (kcal/mol),
  $\Delta \varepsilon_\textrm{HL}^\textrm{gas}$ and $\Delta \varepsilon_\textrm{HL}^\textrm{water}$ are the HOMO-LUMO gaps (eV) in the gas phase and in water respectively,
  Time is the WALL time (min) for one SCF with 160 CPUs,
  in parenthesis the fraction of included integrals is reported.
}
\end{table}

The SCF with the ACE method took 249 min using 160 CPUs on 8 nodes.
With L-ACE(0.001), only 36\% of the integrals are retained in the calculations, 
and the WALL time for the SCF is reduced by almost the same factor (37\%)
with respect to the ACE method.
The error in the hybrid energy correction is of the order of $10^{-4}$
kcal/mol per atom; the errors in the solvation energy is -0.05 kcal/mol;
the error in the HOMO-LUMO gap in water is just 0.0002 eV.
Increasing the threshold of L-ACE the computational times are further reduced:
with L-ACE(0.005) only 16\% of the exchange integrals are included in the SCF,
and the WALL time is 80\% smaller than for ACE.
Errors in the energies and gaps also increase, but still remain small:
the error on the solvation energy is just 0.5626 kcal/mol,
amounting to a relative error of about 0.3\%, while the error on the HOMO-LUMO
gap in water is just 0.002 eV.

\subsection{TiO$_2$ nanoparticles}
As a final example, a TiO$_2$ (anatase) nanoparticle of 465 atoms has been used.
This has the classical bipyramidal shape exposing the (101) and (001) surfaces,
with diameter of 1.5 and 3 nm, in the short and the long dimensions, respectively.
The geometry has been taken from Ref.\cite{mattioli2014}
and is shown in Figure~\ref{fig:all}c, together with the box --
a tetragonal cell with $a=50$ Bohr and $c=70$ Bohr -- in bold black. 
The minimum distance between two atoms belonging to different replicas is 18.36 Bohr and is found between replicas along the $c$ (longest) direction.

The SCF calculations for such systems have been done using the HSE\cite{hse}
functional and the same cutoffs as in the previous Section.
For pure DFT calculations the PBE functional has been used.

\begin{table}[h!]
\footnotesize
\begin{tabular}{llllllll}
 Method        & $\Delta E_\textrm{hyb}$ & $\Delta \varepsilon_{\textrm{HL}}$ & Time         \\\hline
 ACE           & 14.8840               & 4.3692                           & 1877 (100\%) \\ 
 L-ACE (0.001) & 14.8844               & 4.3728                           & 838  (38\%)  \\
 L-ACE (0.002) & 14.8859               & 4.3751                           & 754  (31\%)  \\
 L-ACE (0.003) & 14.8879               & 4.3772                           & 691  (27\%)  \\
 L-ACE (0.004) & 14.8900               & 4.3783                           & 643  (25\%)  \\
 L-ACE (0.005) & 14.8929               & 4.3792                           & 651  (23\%)  \\
 pure DFT      & 0                     & 2.9434                           & 112  (0\%)   \\
\end{tabular}
\caption{\label{tab:tio2}
  Summary of results for the TiO$_2$ cluster (465 atoms).
  Pure DFT calculations with PBE, hybrid calculations with HSE.  
  $\Delta E_\textrm{hyb}$ is the difference between HSE and PBE total energies
  per atom (kcal/mol), $\Delta \varepsilon_\textrm{HL}$ is the HOMO-LUMO gap (eV), 
  Time is the WALL time (min) for one SCF with 288 CPUs,  
  in parenthesis the fraction of included integrals is reported.
}
\end{table}

The system has 2448 electrons, and 1300 KS states have been used in order to compute the HOMO-LUMO gap.
In Table~\ref{tab:tio2} the gaps and the hybrid energy corrections are displayed together with the WALL times for the SCF.
We observe that the SCF with the L-ACE(0.005) method is three times faster than the ACE method,  approaching the performances of the pure DFT. 
The error in the HOMO-LUMO gap is just 0.01 eV.

\section{Conclusions}
\label{sec:conclusions}
We have introduced a new method, L-ACE, for the fast and reliable calculation
of the electronic structure with hybrid functionals and a PW basis set,
effectively combining the ACE method and a localized representation of
molecular orbitals. For the latter, the SCDM  approach has been used. 
We have shown how to integrate L-ACE into the double-loop self-consistent
algorithm of {\sc Quantum ESPRESSO} and how to get rid of 
the unpredictable energy fluctuations present in other algorithms using
truncation.

Compared with other similar strategies based on orbital localization,
\cite{car2009,cervera2009,cervera2007,genovese2015} the present methodology
has the advantage of combining the reduction of the number of evaluations
of the exchange operator, granted by ACE, with the truncation of a large
fraction of terms in each evaluation of the exchange operator, allowed by
localization. Moreover, the algebraic, non-iterative character of SCDM
localization makes the algorithm robust.

In conclusion, L-ACE allows to perform hybrid-functional calculations on
systems that were previously unfeasible or exceedingly costly.
Further work is under way to extend L-ACE to calculations
using k-points
(ACE without localization is already available for this case) 
and to compute each term in the truncated exchange potential via
a real-space solution of the Poisson equation, getting rid of FFTs
altogether.
\newline

\textbf{Acknowledgments}.
This work has been partially funded by the European Union through the
\textsc{MaX} Centre of Excellence (Grant No. 676598). 
We acknowledge Lin Lin, Pietro Delugas and Stefano de Gironcoli for useful discussions about methodological aspects, 
and Giuseppe Mattioli, Loris Ercole and Mariam Rusishvili for providing the geometries for the test cases. 
I.C. is also grateful to Lin Lin for hospitality at the Lawrence Berkeley National Laboratory, where he had very useful discussions about the ACE and SCDM methods.
We acknowledge the CINECA, for the availability of high performance computing resources and support.

\cleardoublepage

\bibliography{biblio}
\bibliographystyle{unsrt}

\end{document}